\documentclass[fleqn,usenatbib]{mnras}

\usepackage[T1]{fontenc}

\DeclareRobustCommand{\VAN}[3]{#2}
\let\VANthebibliography\thebibliography
\def\thebibliography{\DeclareRobustCommand{\VAN}[3]{##3}\VANthebibliography}

\usepackage{newtxtext,newtxmath}

\usepackage{graphicx}	
\usepackage{amsmath}	
\usepackage[export]{adjustbox}

\title{Is binning always sinning? The impact of time-averaging for exoplanet phase curves}

\author[G. Morello et al.]{
Giuseppe Morello,$^{1,2,3}$\thanks{E-mail: gmorello@iac.es}
Achr\`ene Dyrek,$^{4}$
Quentin Changeat,$^{5,6,7}$
\\
$^{1}$Instituto de Astrof\'isica de Canarias (IAC), 38205 La Laguna, Tenerife, Spain\\
$^{2}$Departamento de Astrof\'isica, Universidad de La Laguna (ULL), 38206, La Laguna, Tenerife, Spain\\
$^{3}$INAF- Palermo Astronomical Observatory, Piazza del Parlamento, 1, 90134 Palermo, Italy\\
$^{4}$AIM, CEA, CNRS, Universit\`e Paris-Saclay, Universit\`e Paris Diderot, Sorbonne Paris Cit\`e, F-91191 Gif-sur-Yvette, France\\
$^{5}$Department of Physics and Astronomy, University College London, Gower St., London WC1E 6BT, United Kingdom\\
$^{6}$European Space Agency (ESA), ESA Baltimore Office, 3700 San Martin Drive, Baltimore, MD 21218, USA\\
$^{7}$Space Telescope Science Institute (STScI), 3700 San Martin Drive, Baltimore, MD 21218, USA
}

\date{Accepted XXX. Received YYY; in original form ZZZ}

\pubyear{2022}

\begin{document}
\label{firstpage}
\pagerange{\pageref{firstpage}--\pageref{lastpage}}
\maketitle

\begin{abstract}
We explore how finite integration time or temporal binning can affect the analysis of exoplanet phase-curves. We provide analytical formulae to account for this effect or, if neglected, to estimate the potential biases in the retrieved parameters. As expected, due to their smoother variations over longer time-scales, phase curves can be binned more heavily than transits without causing severe biases. In the simplest case of a sinusoidal phase curve with period $P$, the integration time $\Delta t$ reduces its amplitude by the scaling factor $\text{sinc}{ \left ( \pi \Delta t / P \right ) }$, without altering its phase or shape. We also provide formulae to predict reasonable parameter error bars from phase-curve observations. Our findings are tested with both synthetic and real datasets, including unmodelled astrophysical signals and/or instrumental systematic effects. Tests with the \textit{Spitzer} data show that binning can affect the best-fitting parameters beyond predictions, due to the correction of high-frequency correlated noise. Finally, we summarize key guidelines for speeding up the analysis of exoplanet phase curves without introducing significant biases in the retrieved parameters.
\end{abstract}




\section{Introduction}

The phase curve of an exoplanet describes the total flux received from the star-planet system as a function of the orbital phase. It includes the planetary thermal emission and reflected starlight from the planet, as well as Doppler beaming, ellipsoidal effects, and reflection from the stellar surface (e.g. \citealp{faigler2011,shporer2017}). Phase-curve observations of nearly edge-on systems may exhibit up to two periodic occultations, namely the transit and eclipse of the exoplanet in front of and behind its host star. This geometric configuration is particularly favourable for the characterization of the systems. In fact, the planetary eclipses enable us to separate the stellar and planetary flux contributions. The primary transits are also crucial to determine the size of the emitting/reflecting surface of the planets. We refer to the relevant literature for the detailed geometric description \citep{kipping2010_duration,winn2010}, and light-curve models \citep{louden2018,martin-lagarde2020}.

The optical phase curves of exoplanets are automatically obtained from long-term staring surveys, such as \textit{Kepler} \citep{borucki2010} and the \textit{Transiting Exoplanet Survey Satellite} (\textit{TESS}, \citealp{ricker2014}). Collections of exoplanet phase curves from both \textit{Kepler} and \textit{TESS} missions have been analysed by several authors (e.g. \citealp{esteves2013,esteves2015,wong2020,wong2021}). Dozens of mid-infrared phase-curves have been observed with the \textit{Spitzer}/InfraRed Array Camera (IRAC, \citealp{fazio2004}), mostly on two photometric bands centred at $3.6$ and $4.5 \, \mu\mathrm{m}$ (e.g. \citealp{bell2021,may2022}). Spectroscopic phase curves have also been obtained with the \textit{Hubble Space Telescope} (\textit{HST})/Wide Field Camera 3 (WFC3) at $1.1-1.7 \, \mu\mathrm{m}$ (e.g. \citealp{stevenson2014,kreidberg2018,arcangeli2019,changeat2021,changeat2022,mikal-evans2022}). These observations provide longitudinally resolved information about the exoplanet atmospheres, well beyond the one-dimensional views of the terminator and dayside regions accessible with transit and eclipse spectroscopy \citep{cowan2008,placek2017,parmentier2018}. For these reasons, exoplanet phase-curve observations are expected to play a significant role in the scientific programs of recent and future space missions \citep{bean2018,charnay2022}.

In this paper, we investigate the impact of finite integration time and/or temporal binning on the analysis of exoplanet phase curves. Our work extends an analogous study carried out for the case of transit light curves \citep{kipping2010_binning}. We validate the intuition that binning is less damaging for phase curves, given their smoother variations over longer time-scales, and provide quantitative formulae to predict its effect. In general, the spectrophotometric time series of exoplanetary systems can be taken with various exposure times and/or binned on the fly, leading to integration times from $0.01 \, s$ with a \textit{Spitzer}/IRAC mode up to at least half hour with the \textit{Kepler} and \textit{TESS} long-cadence modes. The main benefits of longer integration times are reduced data storage space and downlink times from satellites (for space missions), but they can also increase the observation efficiency by avoiding many reset gaps. Many data analysers further bins the original time series to speed up the light-curve fits or phase-curve retrievals \citep{mendonca2018data,morello2019,bell2019,changeat2021}. In phase-curve spectroscopy, a few handful of spectra are typically used to sample the full orbital period, the corresponding integration times are above 1 hour \citep{stevenson2014,mendonca2018data,changeat2021,cubillos2021,mikal-evans2022}. On the other hand, too long exposure times can smooth out time-varying signals, causing loss of information and the need to take into account morphological distortions in the model-fits.

The paper is organized as follows. Section \ref{sec:binning} discusses the effects of temporal binning on phase-curve models. Section \ref{sec:bin_maths} introduces mathematical formulae to incorporate the integration interval in time series models. Section \ref{sec:bin_numer} provides numerical estimates for the potential biases due to neglecting the finite integration time in phase-curve models. Section \ref{sec:pc_occ} briefly discusses the different impact of temporal binning to model the transit and eclipse light curves. Section \ref{sec:pc_errorbars} introduces mathematical formulae to estimate the minimum phase-curve parameter error bars and the impact of time-correlated noise, based on tests with synthetic data sets. Section \ref{sec:wasp43b_spitzer} tests the impact of temporal binning on real phase-curve data. Section \ref{sec:discussion} discusses the assumptions made in this study, and Section \ref{sec:charnay_errorbars} compares our formulae for error bars with another one reported in the literature (only for the phase-curve amplitude). Section \ref{sec:conclusions} summarizes the conclusions of this study.

\section{The effects of temporal binning}
\label{sec:binning}

\subsection{Mathematical derivation}
\label{sec:bin_maths}

We assume that the phase curve of an exoplanet is a periodic function of time with period $P$ equal to the orbital period. Consequently, it can be approximated by a Fourier expansion, that can be expressed as the sum of time-lagged harmonics,
\begin{equation}
\label{eqn:pc_sum_cos_lag}
f(t) = \frac{c_0}{2} + \sum_{n=1}^{N} c_n \cos{ \left ( \frac{2\pi n}{P} t - \gamma_n \right ) } ,
\end{equation}
or by using the equivalent sine-cosine form,
\begin{equation}
\label{eqn:pc_sum_sin_cos}
f(t) = \frac{a_0}{2} + \sum_{n=1}^{N} \left [ a_n \cos{ \left ( \frac{2\pi n}{P} t \right )} + b_n \sin{ \left ( \frac{2\pi n}{P} t \right )} \right ].
\end{equation}
Physically motivated models only predicts low-order harmonics with $N \le 3$ \citep{cowan2013,armstrong2015,cowan2017,niraula2022}.

The measured flux is never instantaneous, but is averaged over the detector integration time or over a longer temporal bin set by the user.
We consider an interval of duration $\Delta t$ centred on $t$, so that the measured flux will be
\begin{equation}
F(t) = \frac{1}{\Delta t} \int_{t - \frac{\Delta t}{2}}^{t + \frac{\Delta t}{2}} f(t') dt' .
\end{equation}
By calculating the integral with the expression given in equation (\ref{eqn:pc_sum_cos_lag}), we obtain
\begin{align}
F(t) = & \frac{c_0}{2} + \sum_{n=1}^{N} \frac{1}{\Delta t} \frac{c_n P}{2 \pi n} 
\left [ \sin{ \left ( \frac{2\pi n}{P} \left ( t + \frac{\Delta t}{2} \right ) - \gamma_n \right ) } \right . \nonumber \\
& \left . - \sin{ \left ( \frac{2\pi n}{P} \left ( t - \frac{\Delta t}{2} \right ) - \gamma_n \right ) } \right ] .
\end{align}
Using the Prosthaphaeresis formula, $\sin{a} - \sin{b} = 2 \cos{\left ( \frac{a+b}{2} \right )} \sin{\left ( \frac{a-b}{2} \right )}$, and rearranging, we obtain
\begin{equation}
\label{eqn:pc_integ_final}
F(t) = \frac{c_0}{2} + \sum_{n=1}^{N} c_n \cos{ \left ( \frac{2 \pi n}{P} t - \gamma_n \right ) } \frac{ \sin{ \left ( \frac{2 \pi n}{P} \frac{\Delta t}{2} \right ) } }{ \frac{2 \pi n}{P} \frac{\Delta t}{2} } .
\end{equation}
Comparing equations (\ref{eqn:pc_sum_cos_lag}) and (\ref{eqn:pc_integ_final}), we note that the effect of temporal integration and/or binning is that each harmonic is multiplied by a constant sinc factor, or, equivalently, the coefficients of the Fourier expansion are transformed as follows:
\begin{equation}
\label{eqn:cn_transform}
c_n \to C_n = c_n \text{sinc}{ \left ( \frac{2 \pi n}{P} \frac{\Delta t}{2} \right ) } .
\end{equation}
The same sinc factor applies to get the transformed coefficients of the sine-cosine form.

The nice result from the derivation above is that the binned model time series can be computed analytically, having the same mathematical form as the instantaneous model. Therefore, the correction for the finite time interval can be implemented with very little computational cost in phase-curve calculators, if the underlying models are expressed by sum of harmonic terms, such as \texttt{ExoNoodle} \citep{martin-lagarde2020joss}. Perhaps even better, the correction can be applied a posteriori on the coefficients obtained from the fitting procedure, if required to obtain the correct values of the physical parameters.

\subsection{Numerical estimates}
\label{sec:bin_numer}
As per convention, we adopt the definition of the orbital phase, $\phi = (t-T_0)/P$, where $T_0$ is the epoch of transit, and $\Delta \phi = \Delta t/P$. With this notation, the second-order McLaurin expansion of the sinc factor from equations (\ref{eqn:pc_integ_final}) and (\ref{eqn:cn_transform}) is
\begin{equation}
\label{eqn:sinc_order2}
\text{sinc}{(\pi n \Delta \phi)} \simeq 1 - \frac{\pi^2}{6} n^2 (\Delta \phi)^2 ,
\end{equation}
which is a good approximation if $\Delta \phi \ll 1$, i.e. the temporal bin is much shorter than the orbital period ($\Delta t \ll P$).

If the phase curve consists only of the first harmonic, the temporal binning reduces its amplitude, without altering its phase or shape. We note that even a long bin, such as $\Delta \phi = 1/10$, causes a modest reduction in amplitude, that is, $\sim 1.6$ per cent. This $\Delta \phi = 1/10$ corresponds to a time interval of more than 1 h, even for ultra-short period planets ($P<1 \, d$).
If the phase curve is dominated by the second harmonic, the percentage reduction of its amplitude is four times greater with the same $\Delta \phi$.
The higher order harmonics are more dampened by temporal binning, the percentage reduction of the amplitude being proportional to $n^2$, where $n$ is the order of the harmonic.

In general, phase curves made up of various harmonics are not simply scaled down by temporal binning, but can also change shape. For example, we may expect that the maxima are shifted in phase towards those of the first harmonic, since its weight increases with respect to the other harmonics.
Given that exoplanet phase curves are well represented by low-order harmonics, we anticipate that equation (\ref{eqn:sinc_order2}) provides the correct order of magnitude for the possible changes in morphological parameters. We report some specific examples in Appendix \ref{app:binning_examples}.

\subsection{Phase-curves with occultations}
\label{sec:pc_occ}

The most common and useful studies on exoplanet phase curves are those of near edge-on systems which include transit and eclipse events. These occultations enable us to break several physical degeneracies, e.g. between the planetary size, thermal flux and albedo, as well as the planet-to-star flux ratios \citep{placek2017,parmentier2018}. 

\cite{kipping2010_binning} investigated the effects of binning on the analysis of transit light curves. They found that an integration time of half hour can cause severe biases (if unaccounted for) and/or relatively large uncertainties in the retrieved physical parameters. The analyses of \textit{Kepler}, \textit{TESS} and other synthetic data led to similar findings, also for shorter integration times \citep{howarth2017,morello2018,goldberg2019,huber2022,yang2022}.

It is not surprising that temporal binning affects occultations more than phase-curve modulations, given their different characteristic time-scales and segmented versus smooth variations. In fact, the typical planetary transits and eclipses last a few hours and exhibit steep gradients during the ingress and egress phases, with about per cent variation in flux within tens of minutes. The variations at the contact points are abrupt, so that parametric light-curve models may have discontinuous derivatives (see Appendix \ref{app:discontinuity}). Instead, the typical phase-curve variations are smooth and well below $1$ per cent over the full orbital period of several hours or days.

Given the different criticality of temporal binning during and outside the occultations, we recommend sampling the phase curves with adapted cadences for various portions, in order to save computing time without introducing biases in the retrieved physical parameters.

\begin{figure*}
\includegraphics[width=\textwidth]{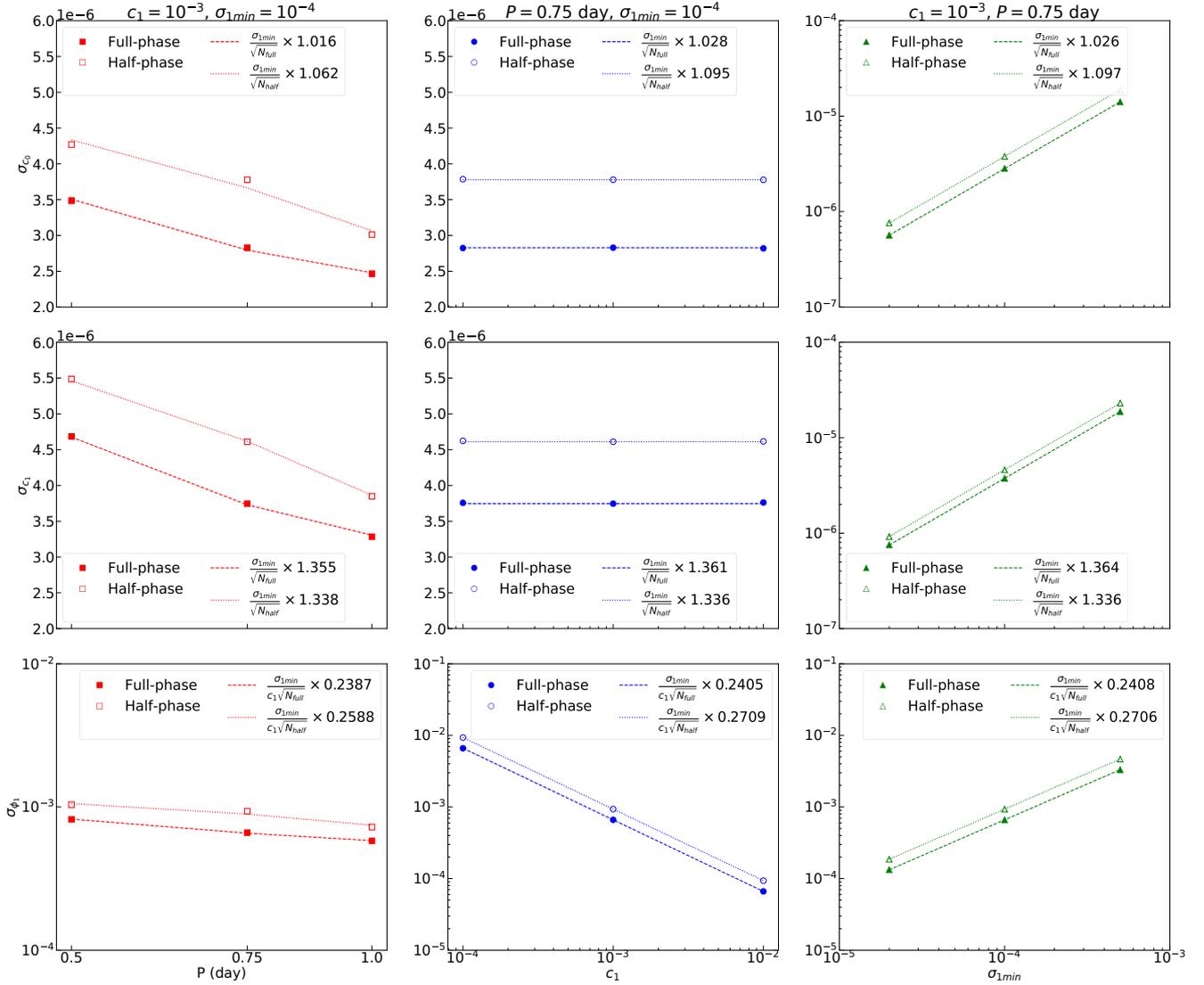}
\caption{Left-hand panels: Phase-curve parameter error bars versus orbital period, based on the model of equation (\ref{eqn:cosinusoid}) with amplitude $c_1=10^{-3}$, and Gaussian noise of $\sigma_{\mathrm{1 min}} = 10^{-4}$. Central panels: Parameter error bars versus $c_1$ for the case with $P=0.75 \, \mathrm{d}$ and $\sigma_{\mathrm{1 min}} = 10^{-4}$. Right-hand panels: Parameter error bars versus $\sigma_{\mathrm{1 min}}$ for the case with $P=0.75 \, \mathrm{d}$ and $c_1=10^{-3}$. Full and empty markers refer to the error bars obtained from the full- and half-phase curves, respectively. The dashed lines denote semianalytical estimates using the formulae specified in the legends.}
\label{fig:errbar_laws}
\end{figure*}

\section{Phase-curve parameter error bars}
\label{sec:pc_errorbars}
In this section, we will introduce simple formulae to estimate the error bars on the phase-curve parameters as a function of the photon noise level, orbital period, and duration of the observation. We will also discuss the effect of time-correlated noise, which can have a stellar or instrumental origin, along with that of time binning on the fitted parameter error bars.

Let us consider one of the simplest phase-curve parametrization, i.e. the shifted cosinusoid
\begin{equation}
\label{eqn:cosinusoid}
f( \phi ) = c_0 + c_1 \cos{[ 2 \pi (\phi - \phi_1 ) ]} .
\end{equation}
For simplicity, we neglect the impact of occultations.
This approximation is analogous to that of ignoring the transit ingress, egress and stellar limb-darkening effects to predict the error bars in transit depth
(e.g. Appendix A of \citealp{morello2019}). In this paper, we do not mathematically derive the formulae for the phase-curve parameter error bars, instead we infer them from multiple fits on synthetic data.

\subsection{Simulations with white noise only}
\label{sec:sim_gauss}
We generated synthetic model light curves with 1 min cadence, based on equation (\ref{eqn:cosinusoid}) with $c_0 = 0$, $\phi_1 = 0$, and the following:
\begin{enumerate}
\item three amplitudes, $c_1 = 10^{-4}$, $10^{-3}$, and $10^{-2}$;
\item three orbital periods, $P = 0.5$, $0.75$, and $1$ d;
\item three values of normalized flux error per frame, $\sigma_{\mathrm{1 min}} = 2 \times 10^{-5}$, $10^{-4}$, and $5 \times 10^{-4}$, the smallest error being similar to the expected noise floor for the \textit{James Webb Space Telescope} \citep{schlawin2020,schlawin2021};
\item two observing windows, $\phi \in [-0.6, 0.6]$ and $\phi \in [-0.2, 0.6]$, corresponding to that of typical full (from eclipse to eclipse) and half (from transit to eclipse) phase-curve observations.
\end{enumerate}
Taking all possible combinations from above, they resulted in 54 configurations. We produced 10 realizations for each configuration by adding or subtracting Gaussian random time series, scaled by the appropriate $\sigma_{\mathrm{1 min}}$, to the noiseless templates. We also created five binned versions of each realization, using binning factors of 1 (unbinned), 10, 30, 60 and 120. In total, we generated 2700 synthetic time series to carry out this experiment.

Then we ran \texttt{EMCEE}\footnote{\url{https://github.com/dfm/emcee}} \citep{emceev3joss} with $20$ walkers and $15 \, 000$ iterations to retrieve the input parameters from all the synthetic time series. The first $5000$ iterations were discarded as burn-in. We adopted wide uniform priors for $c_0$, $c_1$, and $\phi_1$, while keeping the other parameters fixed. We fitted each time series twice, once including and once neglecting the binning correction factor, as given by equation (\ref{eqn:cn_transform}).
The best-fitting parameters are the medians of the respective \texttt{EMCEE} samples (after burn-in) with lower/upper error bars given by the absolute differences between the 16th/84th percentiles and the median.

We computed the best-fitting parameters and error bars for a configuration by averaging those obtained from the individual fits sharing the same baseline model and fitting set-up, i.e. differing only for the specific random noise realization. Note that the same random noise was applied twice, with a `+' or `-' sign, forming pairs of non-independent light-curve realizations. This method provides a shortcut to check for intrinsic bias net of stochastic fluctuations, while also estimating the error bars, instead of using many more independent realizations (see Appendix A of \citealp{martin-lagarde2020}). In fact, the averages between the best-fitting parameters calculated for each pair of light curves are identical and coincide with the input values, except in one case. If the binned light curves are fitted by neglecting the appropriate correction factor, the average of the best-fitting values for $c_1$ coincides with the value predicted by equation (\ref{eqn:cn_transform}). In all cases, the estimated parameter values are identical to those that would be obtained from analogous fits to the noiseless light curves, but the inclusion of noise is crucial to also obtain their error bars. The error bars obtained from the individual fits corresponding to the different noise realizations are very similar, therefore well represented by their averages. The dispersion of individual best-fitting values from analogous configurations is consistent with the estimated error bars.

Having passed all the sanity checks described above, we compared the parameter error bars obtained from the different configurations. Fig. \ref{fig:errbar_laws} shows their main dependencies for the unbinned cases. All parameter error bars are proportional to the inverse of the classical signal-to-noise ratio, $S/N = \sigma_{\mathrm{1 min}}/\sqrt{N}$, where $N$ is the number of points in the light curve. The proportionality factors are $\sim$$1$ for $\sigma_{c_0}$ and $\sim$$1.35$ for $\sigma_{c_1}$. The uncertainty on the phase offset, $\sigma_{\phi_1}$, is also inversely proportional to the amplitude $c_1$. This inverse proportionality is consistent with the fact that the phase of the peak of the sinusoid becomes indeterminate for $c_1 = 0$, as the sinusoid becomes a flat line. We summarize here the formulae to estimate the minimum error bars:
\begin{align}
\label{eqn:sigma_c0}
&\sigma_{c_0} = k_{c_0} \frac{\sigma_{\mathrm{1 min}}}{\sqrt{N}}, \ k_{c_0} \sim 1& \\
\label{eqn:sigma_c1}
&\sigma_{c_1} = k_{c_1} \frac{\sigma_{\mathrm{1 min}}}{\sqrt{N}}, \ k_{c_1} \sim 1.35& \\
\label{eqn:sigma_phi1}
&\sigma_{\phi_1} = k_{\phi_1} \frac{\sigma_{\mathrm{1 min}}}{c_1 \sqrt{N}}, \ k_{\phi_1} \sim 0.25& .
\end{align}

In most cases, binning appears to be harmless, not affecting the parameter error bars. In the most extreme cases, using binning factor $\times 120$ ($\Delta t = 2$\,h) for the shorter period ($P=0.5$\,d , leading to $\Delta \phi = 1/6$), the error bars on $c_1$ and on $\phi_1$ appear to be $\sim 4$ per cent and $6-7$ per cent larger than the corresponding ones without binning. In conclusion, if the phase curve is represented by a single harmonic, it can be binned down to less than 10 time bins without significant loss of information.

\subsection{Simulations with white and time-correlated noise}
\label{sec:sim_gauss_gps}

We created further simulations with additional noise sources other than Gaussian, based on the same model light curves described in Section \ref{sec:sim_gauss}. We used \texttt{celerite}\footnote{\url{https://github.com/dfm/celerite}} \citep{celerite} to model stochastic time series based on a Gaussian process (GP) with a Matern-3/2 kernel,
\begin{equation}
M_{3/2} (\Delta t) \simeq \sigma_{\mathrm{GP}}^{2} \left ( 1 + \frac{\sqrt{3} |\Delta t|}{\rho_{\mathrm{GP}}} \right ) \exp{ \left ( -\frac{\sqrt{3} |\Delta t|}{\rho_{\mathrm{GP}}} \right ) } ,
\end{equation}
where $\Delta t$ denotes the time interval between two data points, $\sigma_{\mathrm{GP}}$ and $\rho_{\mathrm{GP}}$ are characteristic amplitude and time-scale of the stochastic process. Our analysis focused on GPs with short time-scales relative to the orbital period, as they are more likely to be affected by time binning. In particular, we adopted $\sigma_{\mathrm{GP}}=10^{-4}$, and $\rho_{\mathrm{GP}}=3$ and $42 \, \mathrm{min}$, which could mimic stellar granulation and/or instrumental systematic effects \citep{evans2015,chiavassa2017,barros2020}. We computed 10 GP time series for each $\rho_{\mathrm{GP}}$ value, then added a single white noise realization, properly scaled by $\sigma_{\mathrm{1 min}}$, to each of them. These operations resulted in 10 mixed (GP+white) noise time series for each combination of $\sigma_{\mathrm{GP}}=10^{-4}$, $\rho_{\mathrm{GP}}$ and $\sigma_{\mathrm{1 min}}$. We produced 20 light-curve realizations for each parameters configuration by adding or subtracting the mixed noise time series to the noiseless templates. The binned version were also created, as before. In total, we generated 10\,800 synthetic time series to carry out this second experiment.

We followed the same procedure described in Section \ref{sec:sim_gauss} to retrieve the input parameters from the synthetic time series using \texttt{EMCEE}, but including $\sigma_{\mathrm{GP}}$ and $\rho_{\mathrm{GP}}$ as free parameters in the fits. The best-fitting parameter error bars follow the same inverse proportionality trends with the number of data points and phase-curve amplitude, but $\sim$3-10 times larger proportionality factors compared to the case with white noise only. The larger proportionality factors are partly attributable to the greater dimensionality of the fit with GPs. In general, the error bars are not directly proportional to the white noise amplitude, $\sigma_{\mathrm{1 min}}$, but they also depend on both the amplitude and time-scale of correlated noise. Appendix \ref{app:pc_error_gp} presents the detailed results for the tests with white and time-correlated noise.

Some papers with photometric light-curve analyses suggest that temporal binning may be advantageous to reduce the impact of signals with significantly shorter time-scales than those relevant to the astrophysical phenomenon of interest (e.g. \citealp{deming2015,kammer2015}). Our simulations with GPs do not support this claim, the error bars being essentially unaffected by temporal binning. In some cases, the error bars estimated from different binned version of the same light curve can vary by up to $\sim\,20$ per cent in either direction, depending on the specific noise realizations. Fitting the light curves assuming white noise only leads to significantly underestimated error bars, even if adopting a temporal bin much longer than the GP time-scale (e.g. $\Delta t = 120 \, \mathrm{min}$ and $\rho_{\mathrm{GP}} = 3 \, \mathrm{min}$).

\section{Applications}

\begin{figure*}
\includegraphics[width=0.85\textwidth]{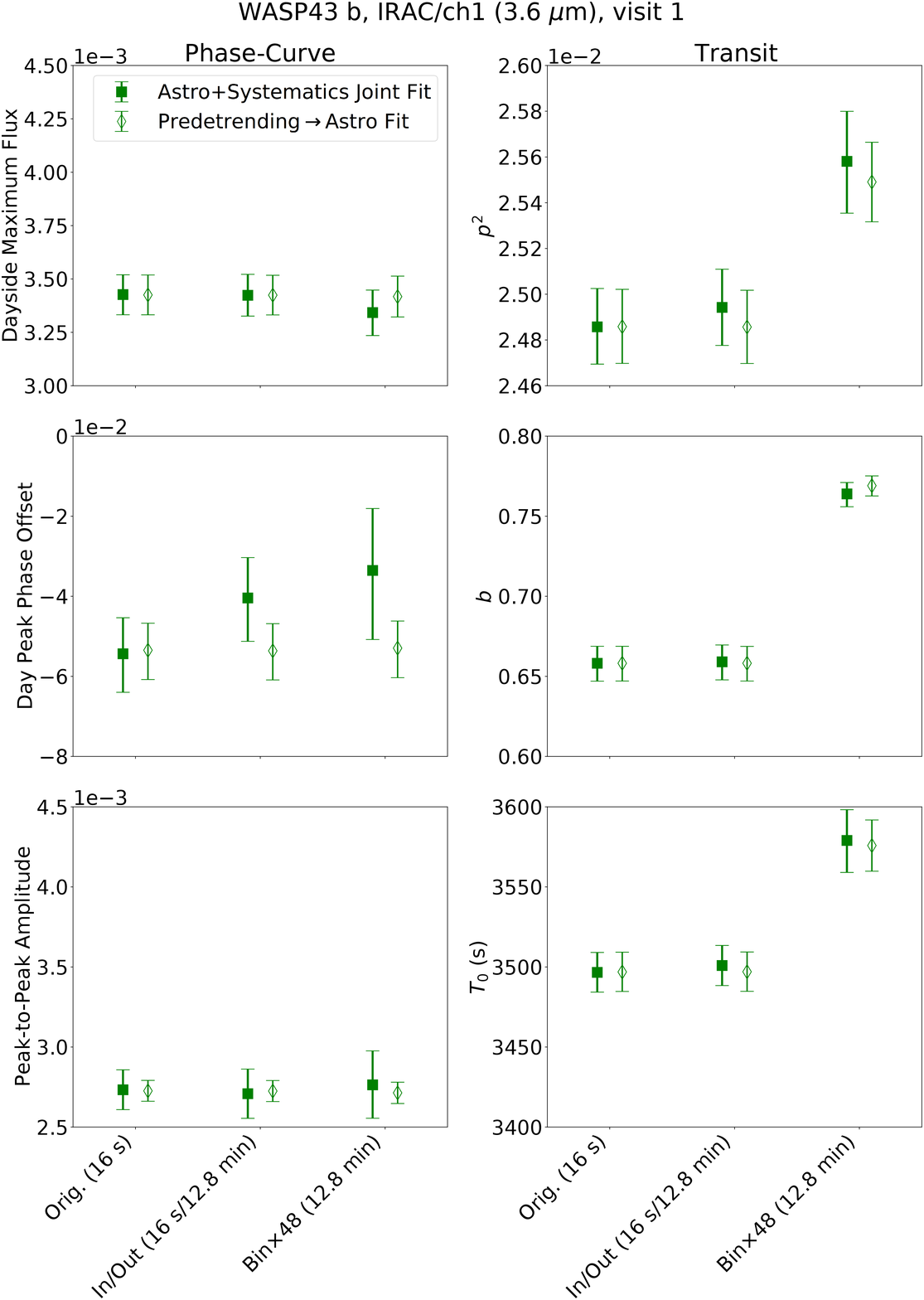}
\caption{Best-fitting phase curve and transit parameters from the \textit{Spitzer}/IRAC $3.6 \, \mu m$ first visit, using three bin configurations and two procedures, as detailed in Section \ref{sec:wasp43b_spitzer}. Using different bin size to sample the occultations and the rest of phase curve is efficient to avoid biases in the retrieved parameters.}
\label{fig:wasp43b_ch1a_results}
\end{figure*}

\begin{figure*}
\includegraphics[width=0.85\textwidth]{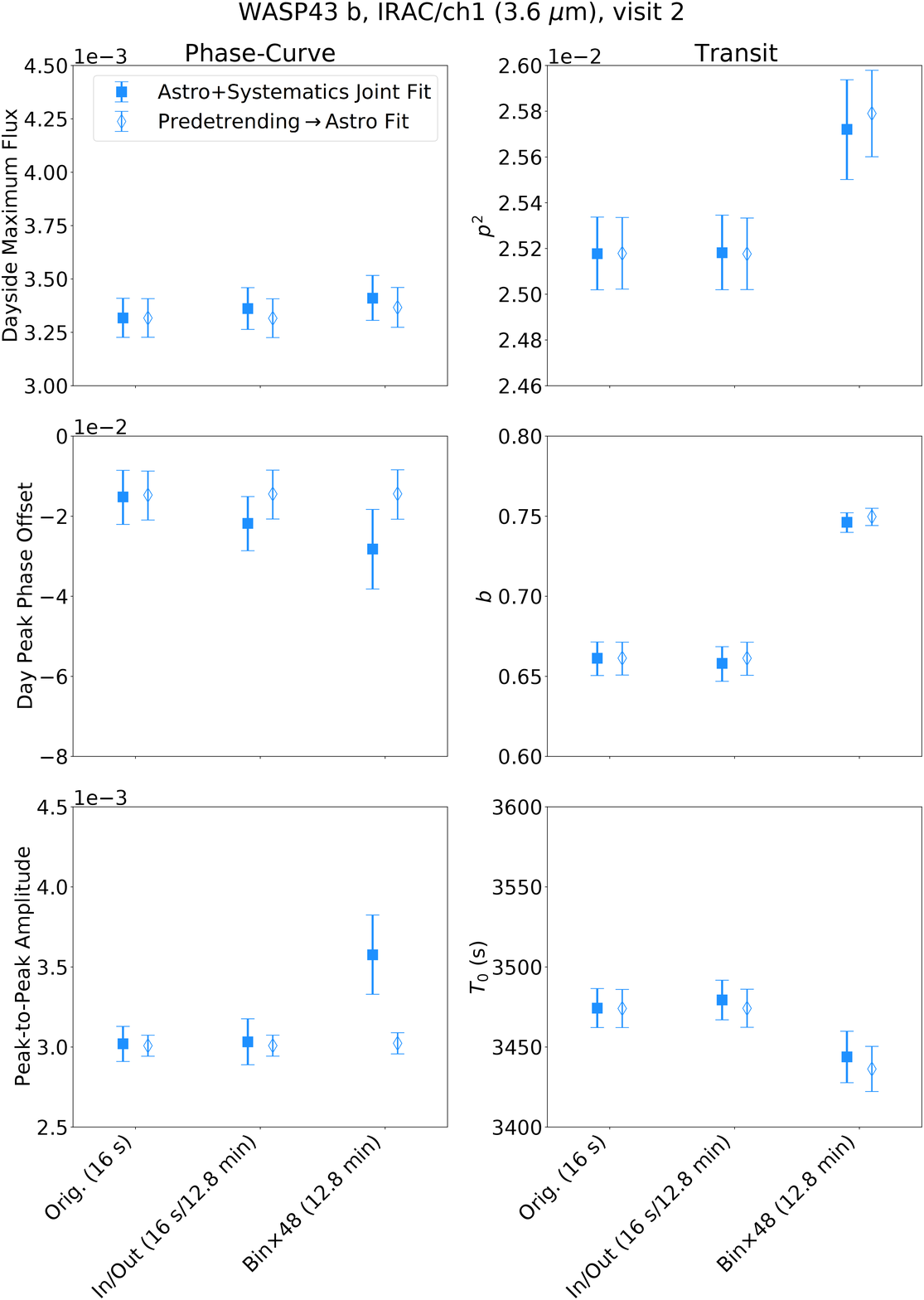}
\caption{Same as Fig. \ref{fig:wasp43b_ch1a_results} from the \textit{Spitzer}/IRAC $3.6 \, \mu m$ second visit.}
\label{fig:wasp43b_ch1b_results}
\end{figure*}

\begin{figure*}
\includegraphics[width=0.85\textwidth]{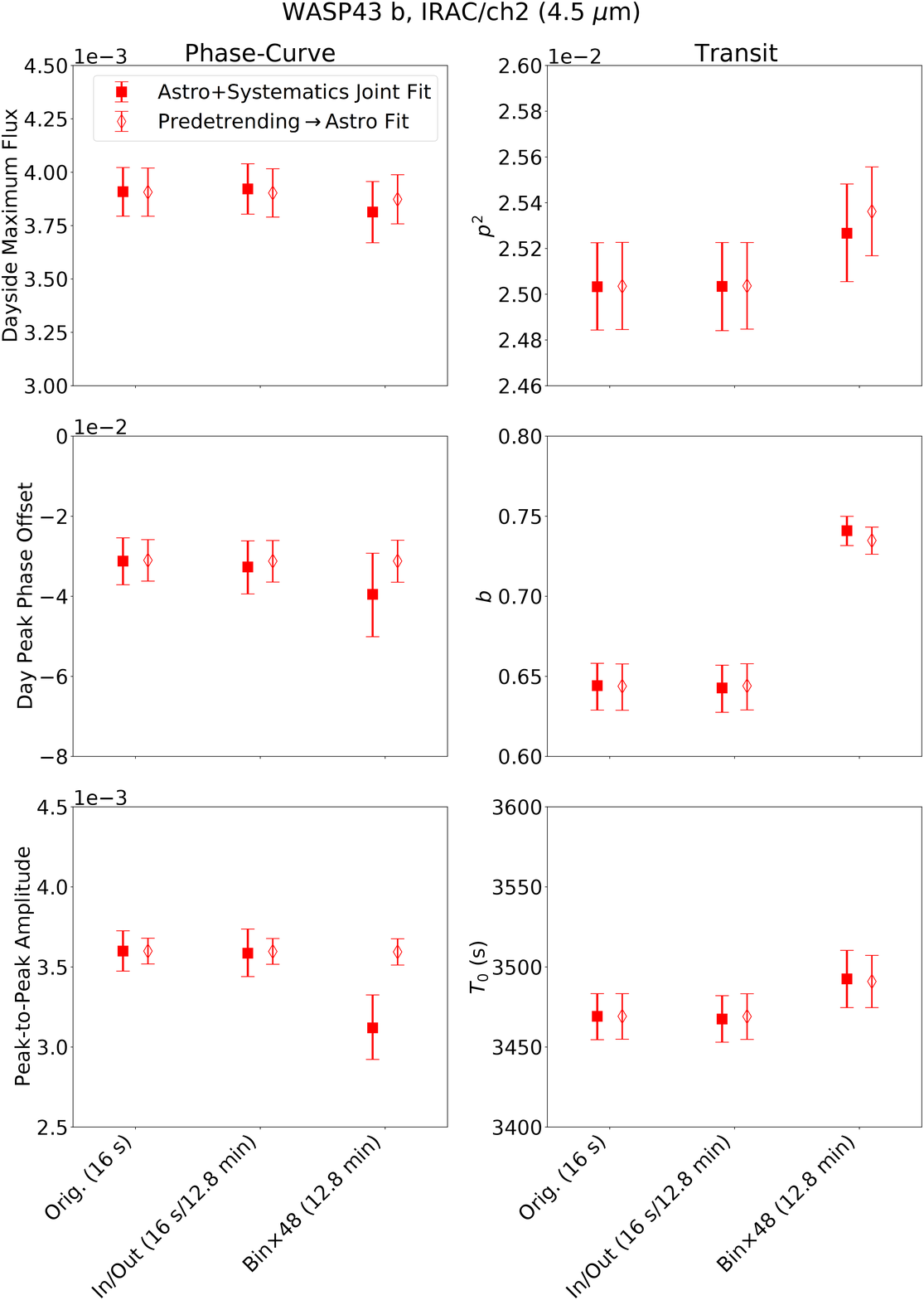}
\caption{Same as Fig. \ref{fig:wasp43b_ch1a_results} from the \textit{Spitzer}/IRAC $4.5 \, \mu m$ visit.}
\label{fig:wasp43b_ch2_results}
\end{figure*}


\subsection{\textit{Spitzer}/IRAC observations of WASP-43 b}
\label{sec:wasp43b_spitzer}

WASP-43 b is a hot Jupiter with an ultra-short period $P=0.813 \, \mathrm{d}$. Three phase-curves of WASP-43 b were observed with the \textit{Spitzer}/IRAC at 3.6 $\mu$m (two visits, program ID 11001) and 4.5 $\mu$m (one visit, program ID 10169). Each visit includes $44 \, 928$ detector images taken with the sub-array mode using $2 \, \mathrm{s}$ frame time, covering one transit and two eclipse events. These observations have been analysed by many authors using various data detrending techniques, leading to some discrepant results \citep{stevenson2017,mendonca2018data,morello2019,may2020,bell2021}. The latest two studies reported that the measured phase-curve amplitudes and offsets may (or may not) vary with temporal binning, especially when using the BLISS algorithm with (or without) an additional Point Spread Function (PSF) width term. The different measurements have led to a debate regarding the heat circulation efficiency of the WASP-43 b atmosphere, and its undergoing physical processes, structure and chemical composition \citep{mendonca2018sim,helling2020,venot2020,changeat2021}.

Here, we evaluate the impact of time binning on the \textit{Spitzer}/IRAC phase curves of WASP-43 b when using the wavelet pixel-ICA data detrending technique \citep{morello2016,morello2019}. This methodology aims to separate the astrophysical signal from instrumental systematic effects by a linear transformation of pixel time series into maximally independent components. For \textit{Spitzer}/IRAC data sets, it is recommended to adopt pixels from a $5 \times 5$ array centered on the target star (see also \citealp{morello2015single}). The raw photometric light curve is fitted with a linear combination of the parametrized astrophysical model and the independent components, excluding (at least) one component that predominantly contains the astrophysical signal.

\cite{morello2019} binned the pixel time series by a factor of eight prior to the ICA transform. The corresponding time interval of $16 \, \mathrm{s}$ was negligible compared to the time-scales of astrophysical interest. Binning was useful to reduce the computational time needed for fitting the light curves. We also find that our ICA algorithm was unable to satisfactorily separate the astrophysical component from the unbinned time series.
In this work, we tested to what extent a much larger binning factor like $384$ (i.e. $8 \times 48$), corresponding to $12.8 \, \mathrm{min}$ interval, affects the results. The number of binned data points is reduced to $117$, which is too low a statistic for robust separation of 25 components via ICA. Instead, we adopted the independent components previously extracted using binning factor of eight (i.e. $16 \, \mathrm{s}$), and binned these components for the light-curve fitting (i.e. $12.8 \, \mathrm{min}$). Apart from the two cases with uniform binning factors, we also fitted the light curves using the smaller bin size around the occultations and the larger one elsewhere, as suggested in Section~\ref{sec:pc_occ}. We applied two fitting procedures for each light-curve and bin configuration. First, we performed a simultaneous fit of the astrophysical model and other independent components, as described by \cite{morello2019}. Second, we subtracted the best-fitting ICA models obtained by \cite{morello2019} from the raw light curves, then fitted the astrophysical model to these detrended light curves. In all cases, the adopted phase-curve parametrization is the sum of two shifted cosinusoids and a constant, 
\begin{equation}
\label{eqn:2cosinusoids}
f( \phi ) = c_0 + c_1 \cos{[ 2 \pi (\phi - \phi_1 ) ]} + c_2 \cos{[ 4 \pi (\phi - \phi_2 ) ]} .
\end{equation}

Figs \ref{fig:wasp43b_ch1a_results}--\ref{fig:wasp43b_ch2_results} compare the results obtained for the three phase curves of WASP-43 b with the bin configurations and fitting procedures described above. In particular, they show three phase-curve parameters, i.e. the maximum flux, peak offset, and peak-to-peak amplitude (computed numerically using equation \ref{eqn:2cosinusoids}), and three transit parameters, i.e. the squared planet-to-star radii ratio ($p^2 = ( R_{\mathrm{p}}/R_* )^2$), impact parameter ($b$), and transit duration of the planet center ($T_0$).

We note a slight dependence of the phase-curve parameters on the binning, when performing the simultaneous fit with the independent components. For the second 3.6 $\mu$m visit, the peak-to-peak amplitude inferred with $12.8 \, \mathrm{min}$ bins is $>1 \sigma$ greater than that obtained with $16 \, \mathrm{s}$ bins. For the 4.5 $\mu$m visit, the peak-to-peak amplitude inferred with the larger bin size is smaller by $>1 \sigma$. Such discrepancies cannot be explained by the pure binning effect, which predicts a negligible underestimation of the amplitude of $\sim 0.02$ per cent, i.e. $<0.01 \sigma$, when using the larger bin size without any correction. The most likely  explanation is that the more heavily binned components fail to correct some high-frequency systematic effects in the \textit{Spitzer} data, also leading to larger error bars by a factor of a few. In fact, when fitting the astrophysical model to pre-detrended light curves, the phase-curve parameters do not show any appreciable variation with different bin configurations.

The transit parameters are significantly biased when using the $12.8 \, \mathrm{min}$ bins without appropriate oversampling, as predicted by \cite{kipping2010_binning}. The bin size adopted to sample the phase curve outside the occultations does not affect the transit parameters.

We conclude that using different bin sizes to model the phase curves during the occultations and outside of them is a suitable strategy to decrease the computational time for light-curve fitting. However, some caution is required to ensure that no time-correlated noise is present with  high-frequency relative to the adopted bin size, otherwise it should be preliminarly corrected. Based on the current analysis, we do not update the phase-curve and transit parameters of WASP-43~b reported by \cite{morello2019}.

\section{Discussion}
\label{sec:discussion}

\subsection{Reliability of the Fourier expansions}
\label{sec:limits_planet}

The derivations made in this work rely on the assumptions that (1) exoplanet phase curves are strictly periodic functions, and (2) low-order harmonics are good enough to approximate the most relevant phenomena. The first assumption is motivated by the expectation that short-period planets are tidally-locked, and it is valid under the hypothesis that their atmospheres have evolved towards an equilibrium state \citep{showman2002}. Several simulations of the atmospheric dynamics of hot Jupiters support a scenario with a static eastward hotspot \citep{cooper2005,showman2008}. Overall, the simulated temperature maps are well approximated by low-order spherical harmonics, which translate into low-order Fourier expansion in time \citep{cowan2013,morris2022}.

Other authors suggested that such simulations are not able to capture the small-scale dynamics and its impact at larger scales, through non-linear interactions \citep{skinner2021}. They predicted the presence of storms with various sizes that cause variability over time-scales longer than the orbital period, leading to quasi-periodic life cycles \citep{Cho_2003,cho2021,skinner2022}. In some cases, the phase curves observed at multiple epochs appear to have significantly different amplitudes and hotspot offsets, although these results may also depend on the data reduction method \citep{armstrong2016,stevenson2017,bell2019,morello2019}.

Even if the Fourier expansion may be inaccurate to model planetary phase curves, the longer term variability should be less affected by temporal binning. Therefore, equation (\ref{eqn:sinc_order2}) would still provide conservative estimates of the order of magnitude of binning effects. Of course, the analytical corrections derived for the harmonics cannot be applied to non-periodic signals.

\subsection{Stellar variability}

Stellar spectrophotometry also exhibits variability at multiple time-scales. Stars often present rotational modulations with typical periods of several days or longer \citep{simpson2010,irwin2011,newton2016,jeffers2018}. They may also present years-long activity cycles \citep{baliunas1985,mathur2014,reinhold2017}. The same considerations concerning the binning effects in presence of longer time-scale signals, as discussed in Section \ref{sec:limits_planet}, are valid regardless of their stellar or planetary origins. Stellar granulation and pulsations may have much shorter time-scales. Their photometric signatures are often, but not always, negligible compared to the planetary phase-curve and/or photon noise limit \citep{kjeldsen1995,barros2020,von_essen2020}. In case of strictly periodic pulsations, the binning effect could be corrected analytically with analogous formulae to those derived in Section \ref{sec:bin_maths}. Sporadic rapid events, such as flares, may introduce subtle biases when using long integration times \citep{gunther2020,jackman2021a,jackman2021b}.

\subsection{Instrumental systematic effects}

Instrumental systematic effects may also introduce signals with various characteristic time-scales. In some contexts, certain data detrending techniques appear to be more/less efficient with binned time series \citep{deming2015,kammer2015,morello2016,may2020,bell2021}. In Section \ref{sec:wasp43b_spitzer}, we explored the impact of different binning factors in the analysis of specific data sets, obtaining slight discrepant parameters ($>1 \, \sigma$) in some cases. A detailed discussion of the instrumental systematic effects in the analysis of exoplanet phase curves is beyond the scope of this paper.

\subsection{Comparison between formulae to estimate error bars}
\label{sec:charnay_errorbars}

\cite{charnay2022} provided a formula to estimate the S/N for phase-curve observations, that we report here:
\begin{equation}
\label{eqn:snr1orbit_charnay}
(S/N)_{\mathrm{1 \, orbit}} = 0.5 \, (S/N)_{\mathrm{1 \, h}}  \, \sqrt{P/2} , 
\end{equation}
where $(S/N)_{\mathrm{1 \, h}}$ is the ratio between the peak-to-peak amplitude and the photometric error over 1 h of integration, i.e.
\begin{equation}
\label{eqn:snr1h_charnay}
(S/N)_{\mathrm{1 \, h}} = \frac{2 c_1}{\sigma_{\mathrm{1 \, h}}} .
\end{equation}
Assuming that $(S/N)_{\mathrm{1 \, orbit}} = c_1 / \sigma_{c_1, \mathrm{1 \, orbit}}$ and replacing $(S/N)_{\mathrm{1 \, h}}$ into equation (\ref{eqn:snr1orbit_charnay}), we obtain
\begin{equation}
\label{eqn:sigma_c1_charnay}
\sigma_{c_1, \mathrm{1 \, orbit}} = \sqrt{2} \frac{ \sigma_{\mathrm{1 \, h}} }{ \sqrt{P} }.
\end{equation}
We note that equation (\ref{eqn:sigma_c1_charnay}) is analogous to equation (\ref{eqn:sigma_c1}), but expressed in terms of 1-h instead of 1-min integrated frames, while $P$ equals the number of integrated frames per orbit. The empirically determined factor $k_{c_1} \sim 1.35$ is similar to $\sqrt{2}$.

\section{Conclusions}
\label{sec:conclusions}

We examined the impact of finite integration time and/or temporal binning in the analysis of exoplanet phase curves. We derived analytical formulae to include these effects in the light-curve models, provided that they can be expressed by the sum of harmonic terms. Additionally, we provide formulae to estimate the minimum phase-curve parameter error bars, depending on the photon noise level, orbital period and duration of the observation.

Owing to smoother variations over longer time-scales, phase curves are less sensitive to temporal binning than transits and eclipses.
We found that bin sizes of up to $\sim$1 h should not significantly affect the retrieved parameters. However, tests with the \textit{Spitzer} data show that, in some cases, the phase-curve parameters vary more than expected with binning due to the correction of high-frequency signals. Hence, some caution is needed to check whether instrumental systematic effects and/or other signals with time-scales shorter than a few hours are present in the data. A significantly smaller bin size and/or appropriate oversampling should be adopted to accurately model the planetary transits and eclipses.

\section*{Acknowledgements}
We thank the referee, David Kipping, for his useful comments that have improved the manuscript. 
G. M. has received funding from the European Union's Horizon 2020 research and innovation programme under the Marie Sk\l{}odowska-Curie grant agreement No. 895525.

\section*{Data Availability}
This research has made use of the NASA/IPAC Infrared Science Archive, which is operated by the Jet Propulsion Laboratory, California Institute of Technology, under contract with the National Aeronautics and Space Administration. The numerical simulations upon which this study is based are too large to archive or to transfer. Instead, we provide all the information needed to replicate the simulations. The original scripts and outputs are available upon request to the authors.
 



\bibliographystyle{mnras}
\bibliography{mybib} 




\appendix

\section{Examples of phase-curve bias versus binning}
\label{app:binning_examples}

\begin{figure*}
\includegraphics[trim={1cm 0cm 2.5cm 0cm}, clip, width=0.49\textwidth]{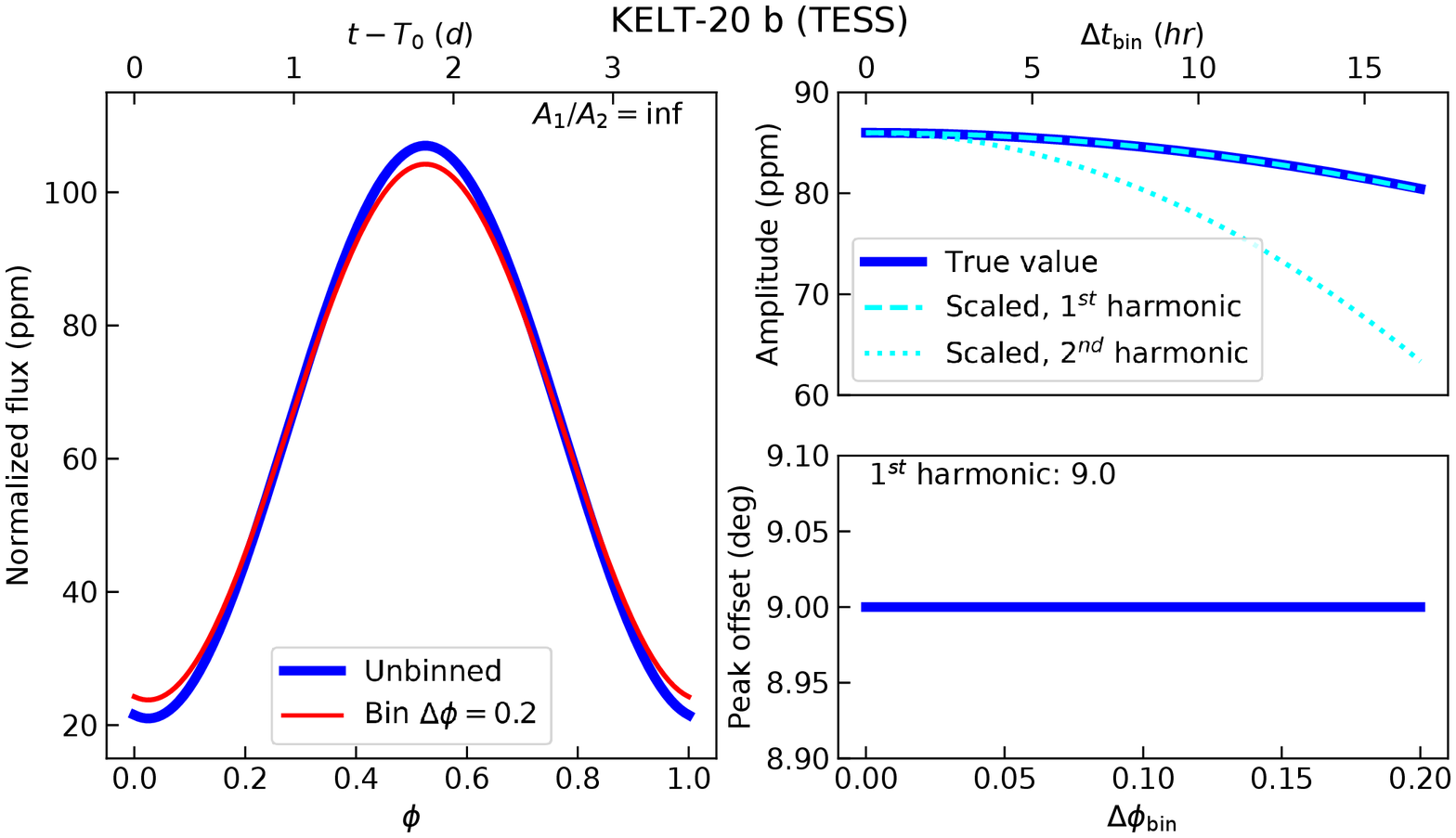}
\includegraphics[trim={1cm 0cm 2.5cm 0cm}, clip, width=0.49\textwidth]{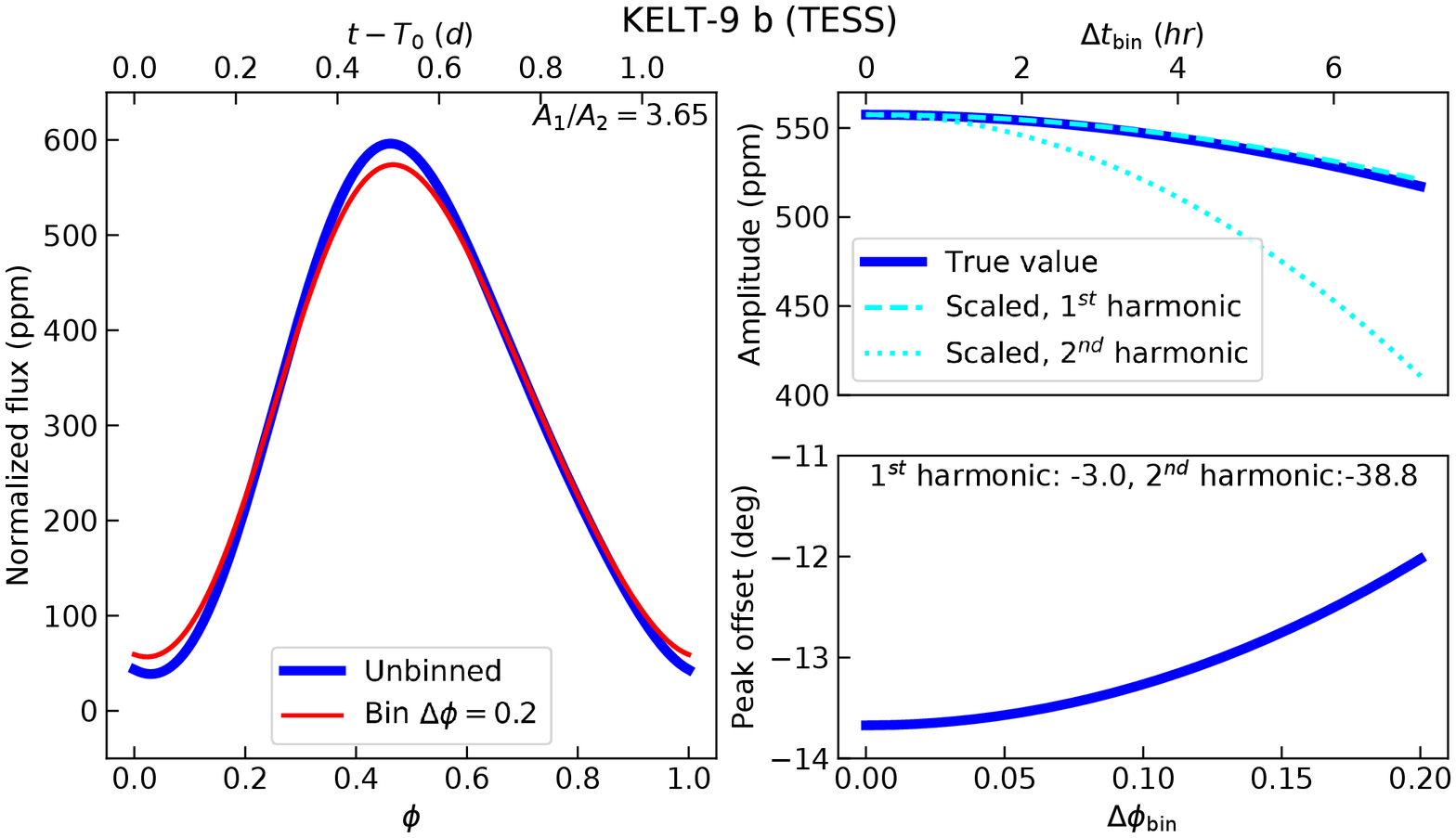}
\includegraphics[trim={1cm 0cm 2.5cm 0cm}, clip, width=0.49\textwidth]{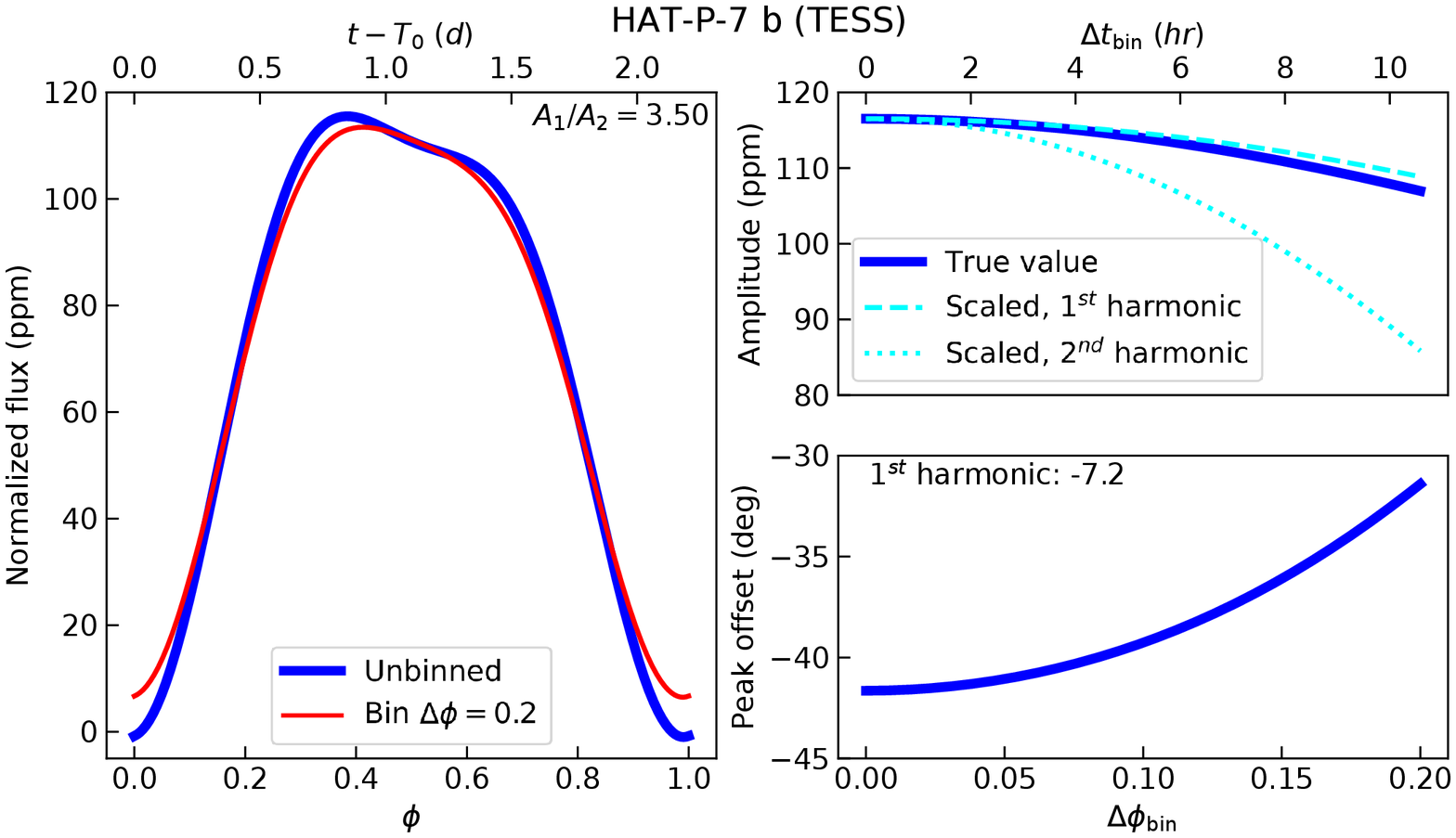}
\includegraphics[trim={1cm 0cm 2.5cm 0cm}, clip, width=0.49\textwidth]{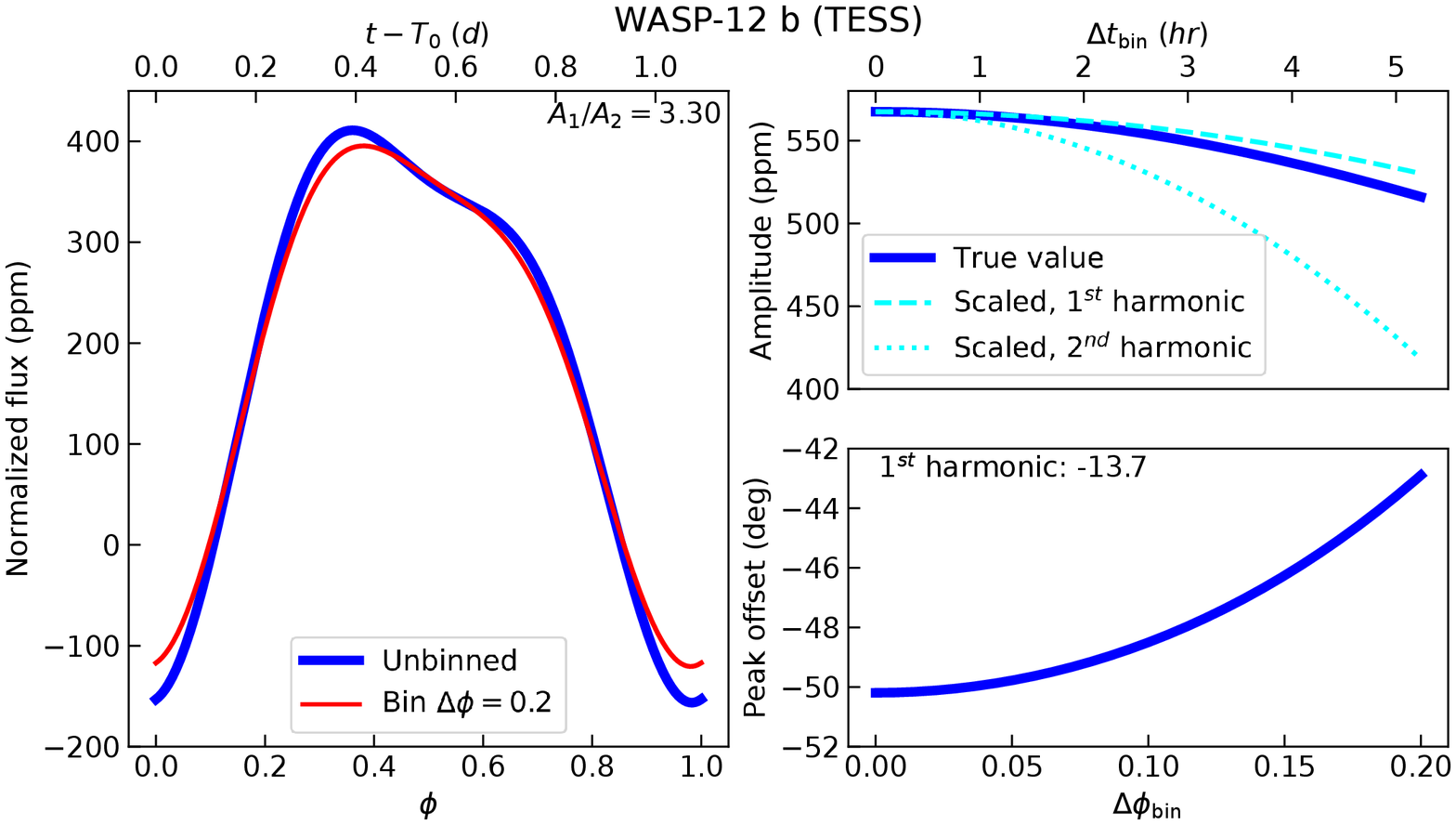}
\includegraphics[trim={1cm 0cm 2.5cm 0cm}, clip, width=0.49\textwidth]{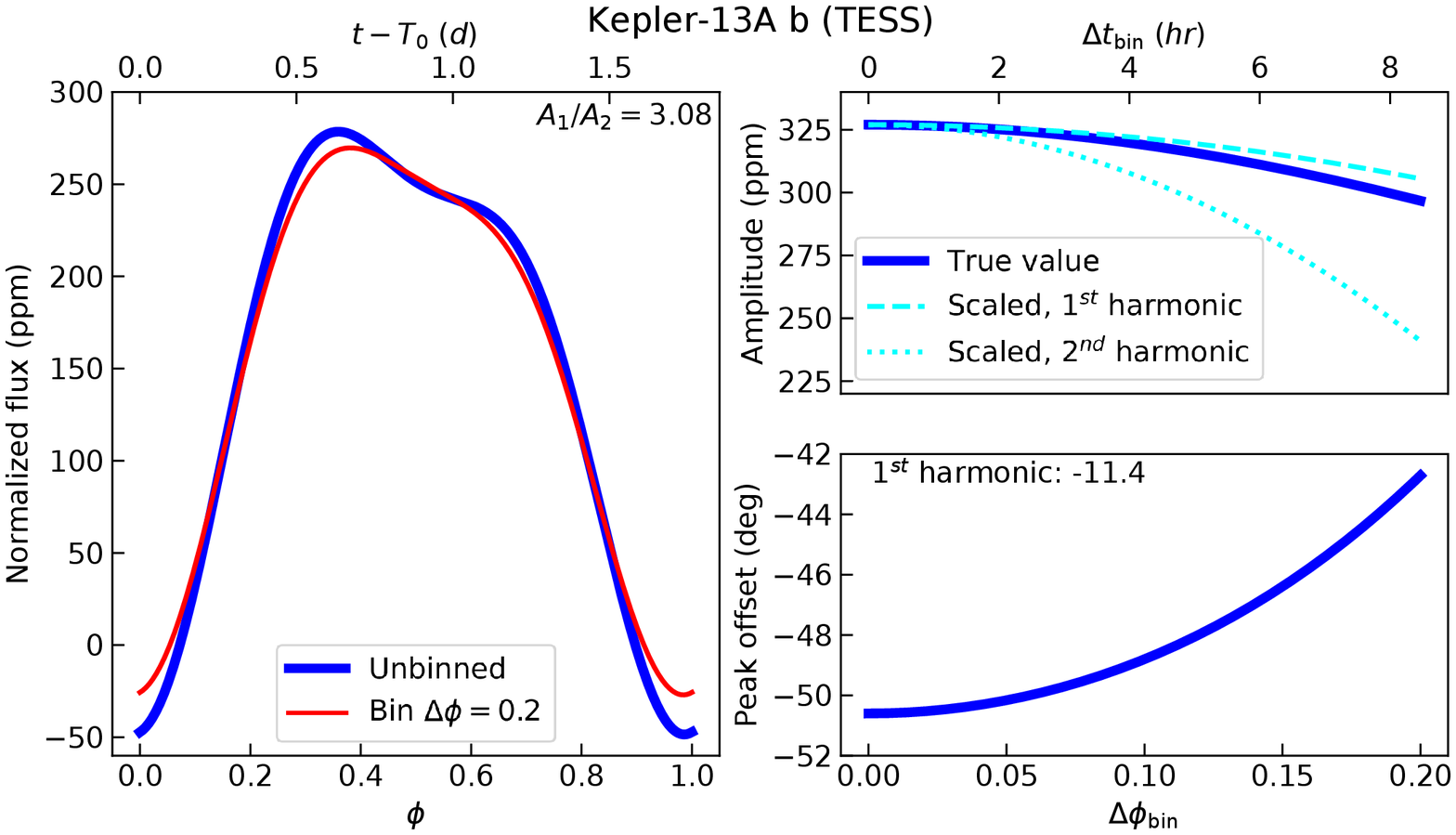}
\includegraphics[trim={1cm 0cm 2.5cm 0cm}, clip, width=0.49\textwidth]{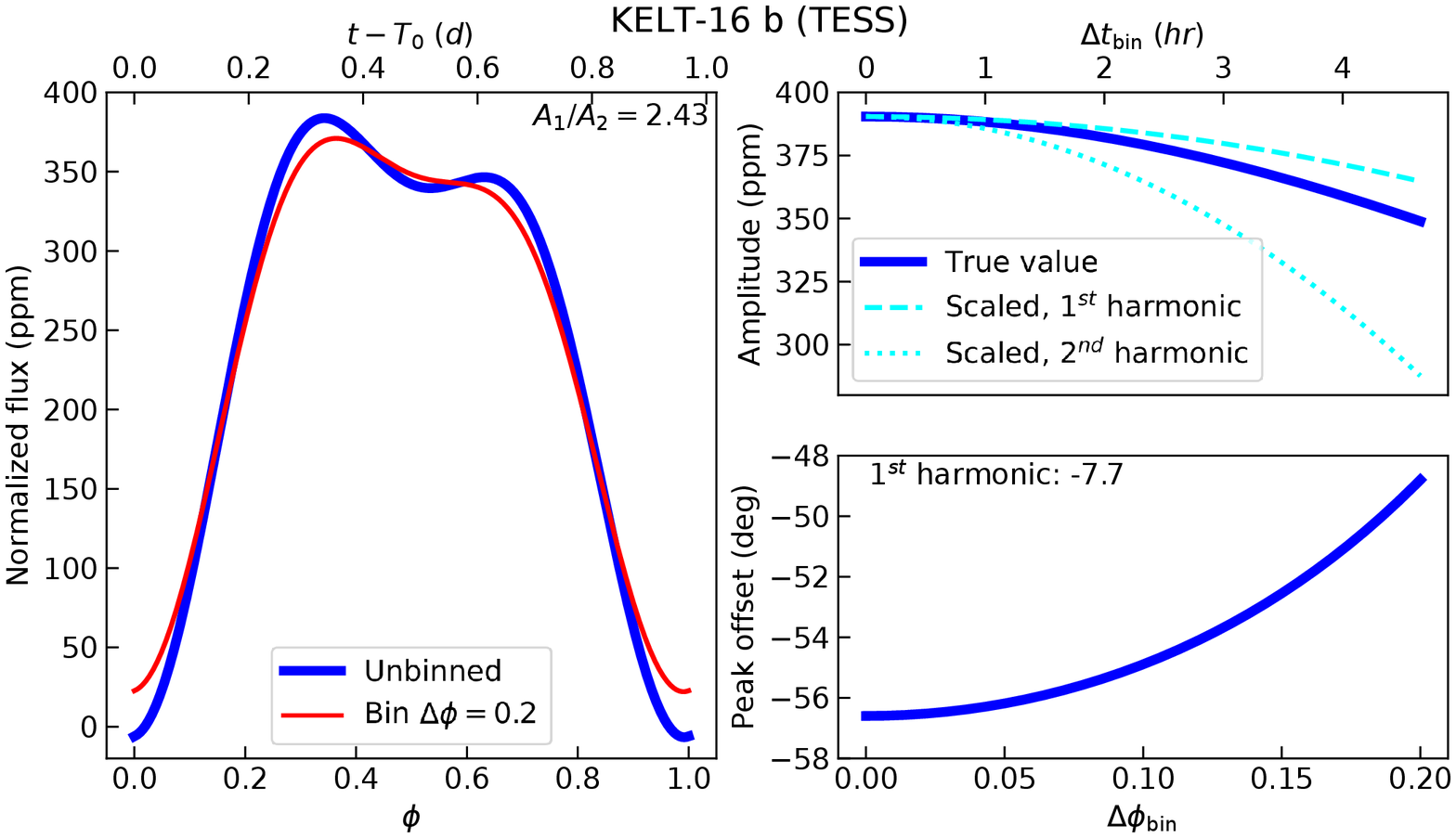}
\includegraphics[trim={1cm 0cm 2.5cm 0cm}, clip, width=0.49\textwidth]{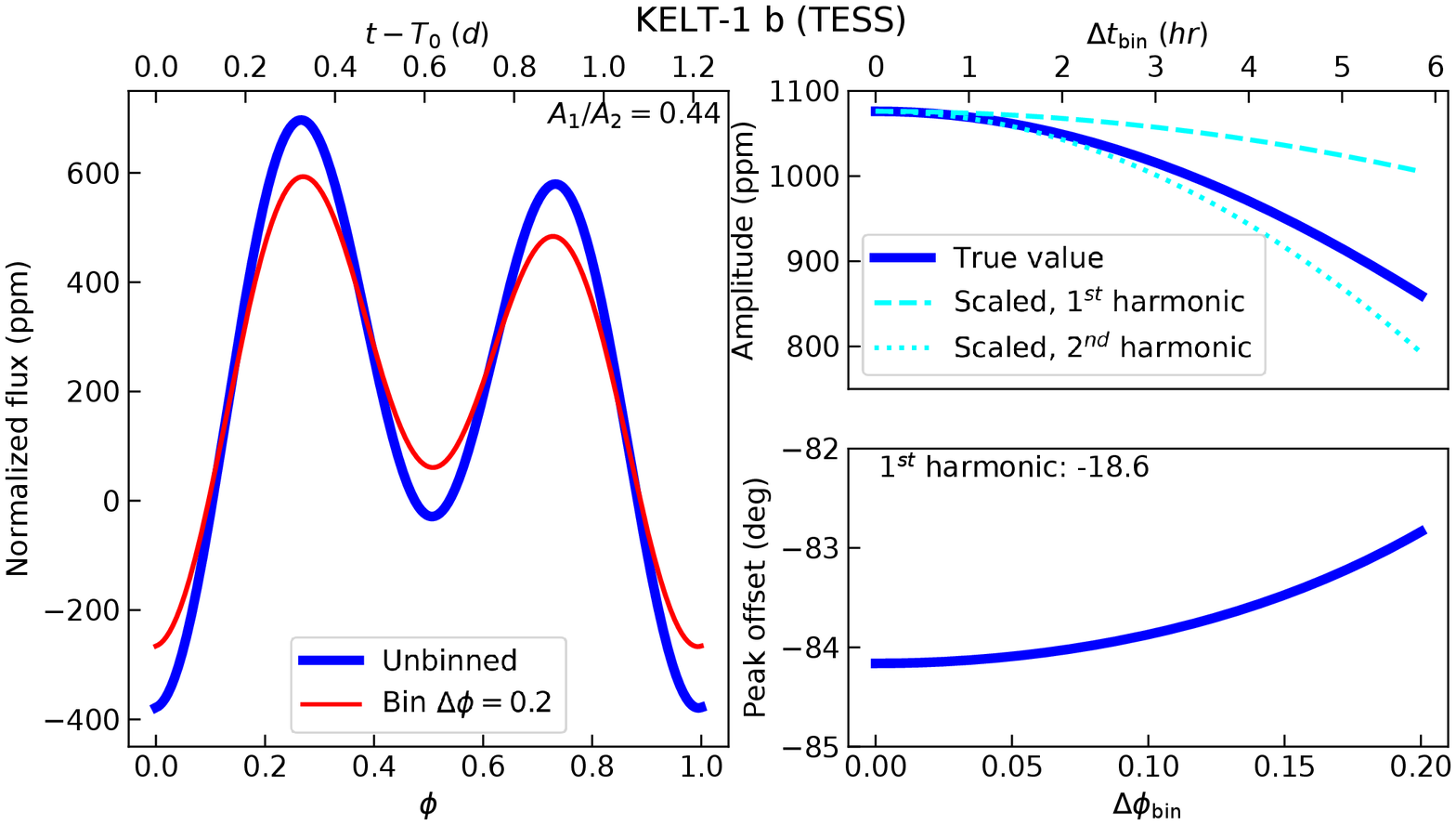}
\caption{\textit{TESS} phase-curve models for various planets as reported by \citet{wong2021}, and relevant parameters versus temporal bin size. Left-hand panels: unbinned light curve (blue) and binned with $\Delta \phi = 0.2$ (red). Top right-hand panels: Apparent peak-to-peak amplitude of the whole phase curve (solid blue line), of the first harmonic (dashed cyan line), and of the second harmonic (dotted cyan line) versus bin size. Bottom right-hand panels: Same as top right-hand panels, but for the peak offset. The peak offset of the second harmonic is always $-90$, except for KELT-9 b.}
\label{fig:wong2021}
\end{figure*}

Fig. \ref{fig:wong2021} shows how temporal binning can affect the peak-to-peak amplitude and offset of the maximum through a range of examples with two harmonics from \cite{wong2021}. In particular, they adopted the following parametrization:
\begin{align}
\label{eqn:wong_planet}
&f_p( \phi ) = \overline{f_p} - A_{\mathrm{atm}} \cos{( \phi + \delta )} ,& \\
\label{eqn:wong_star}
&f_*( \phi ) = 1 - A_{\mathrm{ellip}} \cos{( 2 \phi )} + A_{\mathrm{Dopp}} \sin{ \phi } ,&
\end{align}
where $A_{\mathrm{atm}}$, $\delta$ and $\overline{f_p}$ describe the planet atmospheric term, $A_{\mathrm{ellip}}$ and $A_{\mathrm{Dopp}}$ are the amplitudes of the star ellipsoidal distortion and Doppler boosting. The total phase-curve model is $f( \phi ) = f_p( \phi ) + f_*( \phi )$. Note that the first harmonic is the sum of the planet atmospheric and star Doppler boosting terms. \cite{wong2021} also considered a special parametrization for the atmosphere of KELT-9 b, the hottest giant exoplanet \citep{gaudi2017}:
\begin{equation}
f_p^{\mathrm{KELT-9 \, b}} ( \phi ) = \overline{f_p} - A_{\mathrm{atm}} \cos{( \phi + \delta )} + A_{\mathrm{irrad}} \cos{( 2 [ \phi + \delta_{\mathrm{irrad}} ] )}.
\end{equation}
The bin scaling factor of the peak-to-peak amplitude is intermediate between those of the two harmonics, as from equation (\ref{eqn:cn_transform}). The peak offset is also intermediate between two peaks of the harmonics, and move closer to that of the first harmonic for larger bin sizes.

\section{Phase-curve parameter error bars with GP contamination}
\label{app:pc_error_gp}

\begin{figure*}
\includegraphics[width=\textwidth]{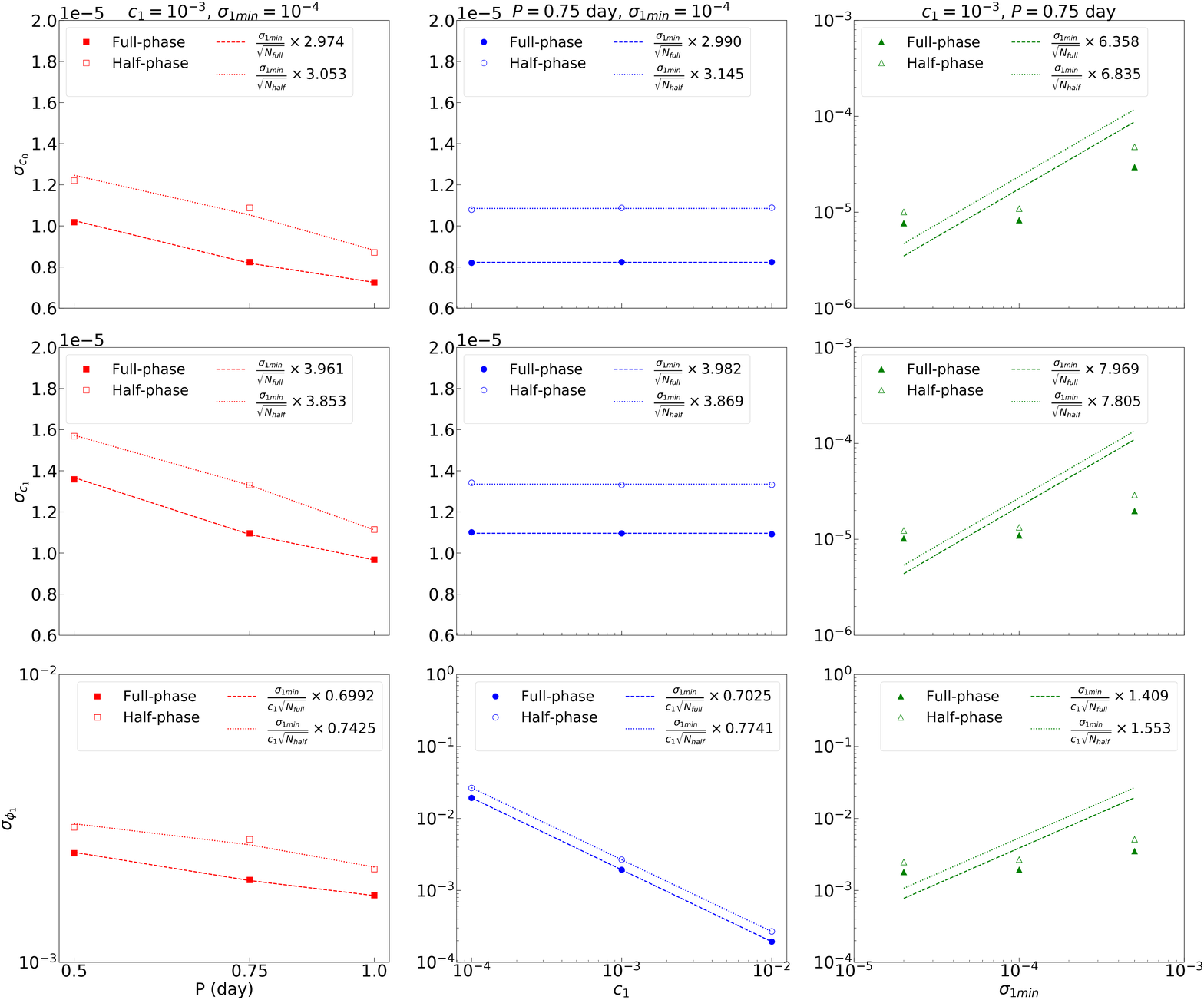}
\caption{Same as Fig. \ref{fig:errbar_laws}, but for the simulations including GPs with $\sigma_{\mathrm{GP}} = 10^{-4}$ and $\rho_{\mathrm{GP}} = 3 \, \mathrm{min}$.}
\label{fig:errbar_laws_gp3min}
\end{figure*}

\begin{figure*}
\includegraphics[width=\textwidth]{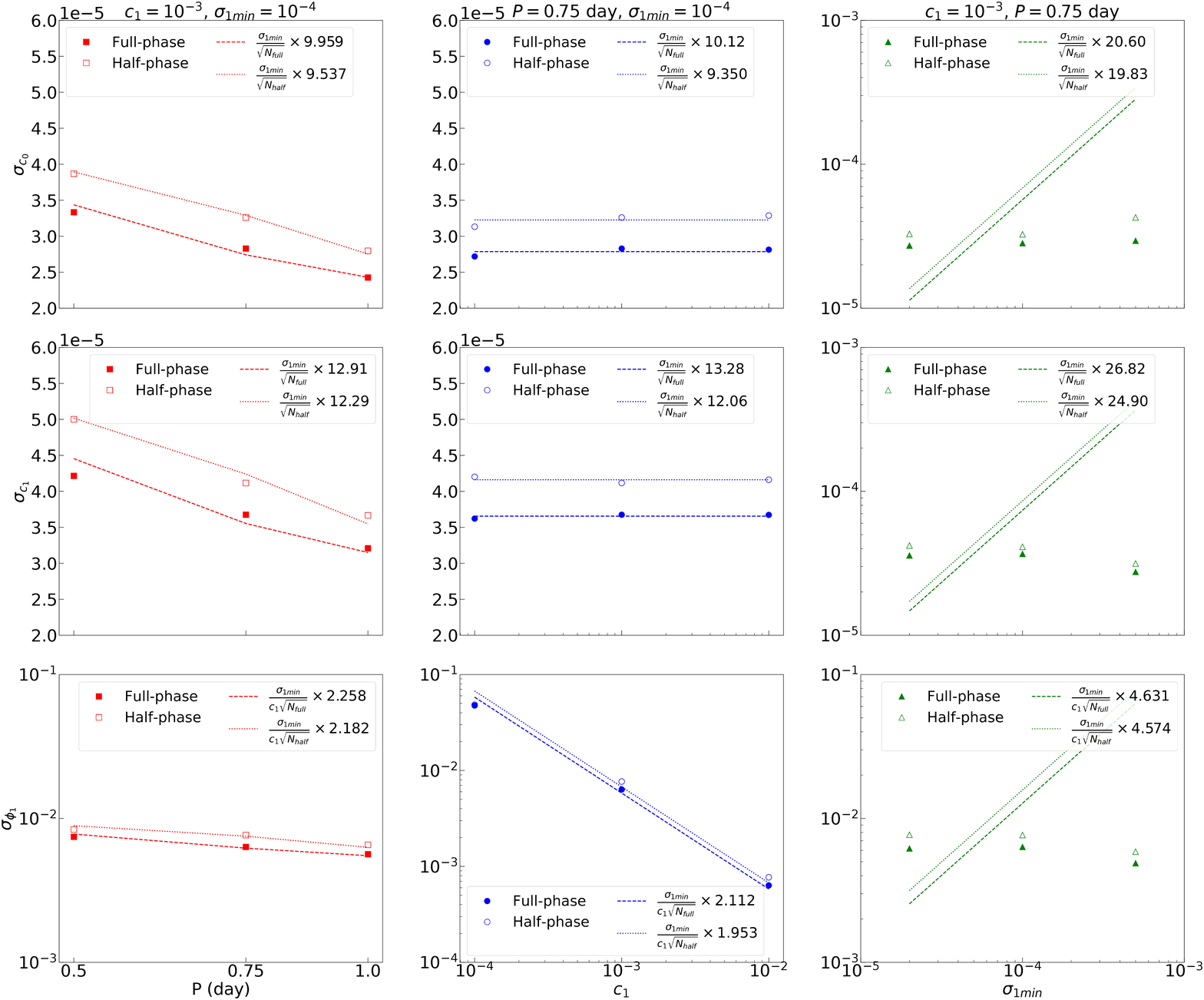}
\caption{Same as Fig. \ref{fig:errbar_laws}, but for the simulations including GPs with $\sigma_{\mathrm{GP}} = 10^{-4}$ and $\rho_{\mathrm{GP}} = 42 \, \mathrm{min}$.}
\label{fig:errbar_laws_gp42min}
\end{figure*}

We show here the phase-curve parameter error bars versus $P$, $c_1$ and $\sigma_{\mathrm{1 \, min}}$ obtained from the simulations with time-correlated noise, as described in Section \ref{sec:sim_gauss_gps}.
Figs \ref{fig:errbar_laws_gp3min} and \ref{fig:errbar_laws_gp42min} report the results obtained with $\rho_{\mathrm{GP}} = 3 \, \mathrm{min}$ and $42 \, \mathrm{min}$, respectively. It clearly emerges that error bars are largely affected by the structure of time-correlated noise. They are significantly larger for the cases with $\rho_{\mathrm{GP}} = 42 \, \mathrm{min}$, which is a more relevant time-scale for the phase-curve modulations (despite being less than one-tenth of the orbital period).

\section{Discontinuities in the derivatives of transit light-curves}
\label{app:discontinuity}

In this Appendix, we review different models used to approximate transit light curves.

\subsection{Box-shaped}
The simplest approximation considers fixed out-of-transit and in-transit fluxes, neglecting the planetary ingress, egress and stellar limb-darkening effects. This function presents two jump discontinuities delimiting the transit event. It is commonly used in $S/N$ calculations \citep{batalha2017,morello2021}, and transit detection algorithms \citep{hippke2019}.

\subsection{Trapezoidal}
The trapezoidal model approximates the planetary ingress and egress by straight lines. The vertices of the trapezoid coincide with the contact points defined by \cite{seager2003}. It provides a convenient analytical formula for mathematical explorations \citep{carter2008,kipping2016}, and it is also used in transit detection algorithms \citep{hippke2019,li2019}. The derivative of the trapezoidal model is a piecewise constant function with a negative/positive value during the ingress/egress, and null during the full transit and out of transit.

\subsection{Uniform source}
If the star is represented by an uniform emitting disc, the decrease in flux during transit is proportional to the fraction of occulted stellar area. Alike the trapezoidal model, the flux is constant (null derivative) between the second and third contact points and out of transit. Following the notation of \cite{mandel2002}, the normalized flux is a function $F(p,z) = 1 - \Lambda(p, z)$, where $p$ is the planet-to-star radii ratio and $z$ is the projected star-planet distance  in units of the stellar radius ($z$). Here, $\Lambda(p, z)$ is the fraction of stellar disc occulted by the planet:
\begin{equation}
\label{eqn:lambda_transit}
\Lambda(p, z) = \begin{cases}
0 & z>1+p \\
\frac{1}{\pi} \left [ p^2 k_0 + k_1 - \sqrt{ \frac{4z^2 - (1 + z^2 - p^2)}{4} } \right ] & |1-p| < z \le 1+p \\
p^2 & z \le 1-p \\
1 & z \le p-1
\end{cases}
\end{equation}
being
\begin{equation}
k_0 = \arccos{\left ( \frac{p^2 + z^2 -1}{2pz} \right )}
\end{equation}
and 
\begin{equation}
k_1 = \arccos{\left ( \frac{1 - p^2 + z^2}{2z} \right )}.
\end{equation}
The derivative of equation (\ref{eqn:lambda_transit}) is
\begin{multline}
\frac{ \partial \Lambda }{ \partial z }(p,z) = \\
\begin{cases}
\frac{1}{\pi} \frac{\sqrt{(1+z+p)(1+z-p)(1-z+p)(1-z-p)}}{z} & |1-p| < z \le 1+p \\
0 & \text{elsewhere}
\end{cases}.
\end{multline}
Note that
\begin{equation}
\lim_{z \to |1-p|^+} \frac{ \partial \Lambda }{ \partial z }(p,z) = \lim_{z \to (1+p)^-} \frac{ \partial \Lambda }{ \partial z }(p,z) = 0 ,
\end{equation}
i.e. $\frac{ \partial \Lambda }{ \partial z }(p,z)$ is a continuous function, even at the contact points.
The second derivative is
\begin{multline}
\frac{ \partial^2 \Lambda }{ \partial z^2 }(p,z) = \\
\begin{cases}
\frac{1}{\pi} \frac{z^4 - (p^2 - 1)^2}{z^2 \sqrt{(1+z+p)(1+z-p)(1-z+p)(1-z-p)}} & |1-p| < z \le 1+p \\
0 & \text{elsewhere}
\end{cases}.
\end{multline}
Note that
\begin{equation}
-\lim_{z \to |1-p|^+} \frac{ \partial^2 \Lambda }{ \partial z^2 }(p,z) = \lim_{z \to (1+p)^-} \frac{ \partial^2 \Lambda }{ \partial z^2 }(p,z) = + \infty ,
\end{equation}
i.e. $\frac{ \partial^2 \Lambda }{ \partial z^2 }(p,z)$ is discontinuous at the contact points.

\subsection{Limb-darkened source}
In many applications, the radial decrease in specific intensity on the stellar disc is parametrized by a limb-darkening law \citep{claret2000,howarth2011,morello2017}. Analytical expressions of the flux derivatives for some laws have been published in previous studies \citep{pal2008,agol2020}. 
Limb-darkening can cause the transit light curves to appear rather U-shaped instead of almost trapezoidal. Nevertheless, the second- or higher order flux derivatives are discontinuous at the contact points.


\bsp	
\label{lastpage}
\end{document}